# Fault Current Limiter Dynamic Voltage Restorer (FCL-DVR) with Reduced Number of Components

Pedram Ghavidel, Masoud Farhadi, *Student Member, IEEE,* Morteza Dabbaghjamanesh, *Senior Member, IEEE,* Alireza Jolfaei, *Senior Member, IEEE,* Mehran Sabahi

*Abstract*— In this paper, a new dual function fault current limiter-dynamic voltage restorer (FCL-DVR) topology is proposed. The proposed structure, in addition to performing routine FCL tasks, can be used to improve the voltage quality of point of common coupling (PCC). A salient feature of this FCL-DVR is its reduced number of semiconductor switches and gate drive and control circuit components. Perhaps, variety structures of FCL-DVR have been proposed but most distinctive feature of proposed structure is lower power loss. The operation modes and the control strategy of the FCL have been presented and studied. In addition, the proposed structure has been compared with other structures to prove the efficiency of proposed structure. The simulation results as well as experimental outcomes from a laboratory scaled-down prototype are provided, which prove the efficiency and feasibility of the proposed structure.

Nomenclature

| | |
|---|---|
| $C_1$ | Output filter capacitance (F) |
| $L_m$ | Magnetizing inductance (mH) |
| $L_r$ | Leakage inductance (mH) |
| $U_{comp}$ | Series injected voltage (V) |
| $t_f$ | Time of fault occurrence (S) |
| $L_t$ | Total of inductance of source, transmission line and transformer leakage inductance (mH) |
| $R_t$ | Total resistance of source and transmission line (Ω) |
| $L_L$ | Load inductance (mH) |
| $R_l$ | Load resistance (Ω) |
| $\omega_0$ | Natural frequency of the network (Hz) |
| $Z_{FCL}$ | Impedance of FCL |
| $Z_{Line}$ | Impedance of transmission line |
| $\alpha$ | Percentage of line voltage variation after fault |
| $V_{Lr}$ | Voltage drop in the transformer leakage inductance |
| $a$ | Turns ratio of the transformer |
| $\lambda$ | Ratio of injected voltage to line voltage |
| $P_{core}$ | Iron loss of transformer (W) |
| $P_{cu}$ | Copper loss of transformer (W) |
| $P_{igbt}$ | IGBT losses (W) |
| $R_{T1}$ | Resistance of the primary winding (Ω) |
| $R_{T2}$ | Resistance of secondary winding (Ω) |
| $V_{CES}$ | Forward blocking voltage of IGBT (V) |

P. Ghavidel and M. Sabahi are with the Faculty of Electrical and Computer Engineering, University of Tabriz, Tabriz 51666, Iran (e-mail: p.ghavidel93@ms.tabrizu.ac.ir; sabahi@tabrizu.ac.ir).
M. Farhadi and M. Dabbaghjamanesh are with the Erik Jonsson School of Engineering, The University of Texas at Dallas, Richardson, TX 75080 USA (e-mail: Masoud.Farhadi@utdallas.edu; mortezadabba@utdallas.edu).
A. Jolfaei is with the School of Information and Communication Technology, Griffith University, Gold Coast, QLD 4222, Australia (e-mail: alireza.jolfaei@griffithuni.edu.au).
Color versions of one or more of the figures in this article are available online at http://ieeexplore.ieee.org.

## I. INTRODUCTION

THE promotion of modern electric power system and their interconnections have led to the high fault currents levels that may go beyond of the maximum short-circuit ratings of the switchgears in some points of the grid. Despite advantages of modern networks, large power networks exhibit some problematic issues such as high complexity, lower security, and high-level fault currents [1], [2]. Furthermore, these power networks suffer from the probable decrease in the quality of the load voltage due to the nonlinear loads and high penetration of renewable energy systems. On the other hand, in a competitive electricity market, the voltage improvement is very crucial, and companies must improve the quality of the power provided to the consumer in order to gain more share of the electricity market. Also, by increasing the number of sensitive loads, importance of the power quality becomes a priority in the operation of power delivery system.

Among different types of disturbances that can occur in the power system, such as voltage sags, voltage swells, imbalances, frequency deviations, flicker, and harmonics interruption, voltage sag is known to produce the most devastating impact on the loads [3], [4]. Voltage sag is mainly a result of short-circuits in the grid. Studies have shown that 92% of all disturbances in the industrial distribution systems are voltage sags, transients, and momentary interruptions [5], [6]. The importance of the power quality has prompted installation of power conditioning equipment to mitigate voltage disturbances. Many solutions are available to improve power quality [7]. Among these solutions, dynamic voltage restorer is the most deployed solution to compensate for voltage sags [8]. DVR consists of a voltage source inverter (VSI) to inject voltage in series with the line, injection transformer, and a dc link for voltage sag interval. The DVR injects voltage to compensate the sag and as a result, the load voltage remains almost constant.

Today, with increasing number of faults and increase in the current amplitude at the time of the occurrence of the fault, the need for installation and use of fault current limiting to protect the network becomes a necessity. An efficient design of FCL should have a low number of elements and a simple control circuitry. In addition, it must limit the fault current efficiently and improve the voltage quality perfectly. From costs perspective, a structure with less number of elements is usually economical. Recently, different types of FCLs are presented with different features [9]-[11], [27]. Whereas the need for a structure that can limit the fault currents is investigated in literature, a simultaneous voltage sag compensation or harmonics compensation is studied just in few literatures such as [9]-[12]. In [9], a novel structure is proposed with both fault current limiting ability and dynamic voltage restorer functionality. Proper control strategy in this structure enables the low power loss during short circuit limiting and voltage restoration. This can reduce the current of power supply during fault current limiting. In [10], a dynamic voltage conditioner is analyzed to find the fault current limiting availability. Therefore, within the proper range, the opening time of upstream circuit breaker can be guaranteed. In [11], a virtual impedance based FCL is proposed to be used in downstream DVRs. This structure can insert high series impedance to line to reduce the fault current. In [12], active solutions for embedding fault current limiters into dynamic voltage conditioner are investigated. An auxiliary controller for downstream FCL in a radial distribution line by means of a DVR is presented in [8]. This control scheme has two inner and outer loops for providing damping for the transients caused by the DVR filter and controlling the injected voltage magnitude and phase angle of the faulty phase to interrupt the fault current and restore the PCC's voltage, respectively. In [13]-[15], virtual current-limiting impedance technique is used to create high impedance during fault. To this end, a virtual RL impedance upon sensing of a voltage sag will be insert to the line. However, this can cause circulating active power through the series and shunt converter. To alleviate this issue, the flux-charge model scheme which only insert a virtual inductor in series with line, is recommended in the [13]–[15]. In comparison with RL feedforward scheme, this method can limit the line current by absorbing of nearly zero active power by the converter. Unlike [13]–[15] which recommend fixed virtual impedance insertion, in [16] the FCL–DVR behaves as a virtual inductance with a variable value. Since inductance is utilized to avoid large swings in this scheme, the real power absorption is inevitable, and the DC-link capacitor will be aged quickly. Another scheme to achieve both DVR and FCL functionalities is to use the filter inductor as a part of the *LC* filter under normal operation conditions, as well as an impedance to limit the fault current under fault conditions [17]–[19]. In [20], a FCL–DVR is proposed by combining various topologies. In the fault condition, the controller deactivates the converter in faulty phase and activates the DC-link semiconductor switches, therefore, an equivalent bridge-type current limiter will be establish to restrict the fault current [20]–[21].

In [8], the voltage sag in the presence of superconducting FCL (SFCL) is investigated. The superconductor based FCLs have a low or no power loss during normal and fault conditions, so they are attractive in industry [26]-[27]. However, in recent years due to the high technology demands and higher costs of the superconductive materials, application of non-superconductor FCLs is recommended [1], [9]. Apart from the great diversity of FCL's structures, an effective FCL should have the following characteristics:

• Limit the fault current in the shortest possible time to prevent stability issues
• Exhibit low losses especially during the normal operation of power network
• Should have no impact on the network operation during normal operation
• Minimal number of elements
• Low price

In this paper, the proposed structure can compensate the voltage sag as well as limit the fault current effectively while having a minimum number of elements. To this end, the proposed FCL-DVR uses an active impedance and voltage injection strategy. The main features of this structure are as follows:

• The simplicity of the power circuit topology
• Fewer numbers of elements compared with most of the existing structures of FCL-DVR, which reduces the construction costs. Because of the necessity of FCL for effectively limit the fault current and given the widespread use of the FCL in power system, this feature is one of the major contributions of the proposed structure. In addition, it should be noted that based on field experience, power switches are the most vulnerable components in the power electronics systems [28]–[30]. Therefore, the proposed FCL-DVR would be a promising topology from the reliability point of view.
• In addition to limiting the fault current, the proposed structure can also improve voltage quality by compensation of voltage sag like a general dynamic voltage restorer (DVR). The fault location is at PCC and FCL–DVR is supposed to control the voltage at the point of common coupling (PCC).

## II. PROPOSED FCL-DVR

Generally, due to the configuration of both FCLs and DVRs, they can only perform a singular function, that is either limiting the fault current or compensating the voltage sags respectively. A structure which can perform both functions would be very attractive to the industry because it will have the ability to protect sensitive loads from utility's most common voltage quality problem such as voltage sags, while at the same time protecting the utility electrical installations from damaging fault currents. This will make it possible to avoid catastrophic shutdowns due to breakdown of power switches caused by high fault currents. The proposed structure considers both abovementioned features. Based on Table I, the number of elements used in the proposed structure has decreased. Furthermore, the implementation of the proposed structure is much simpler compared to structures used in [31], [32]. The proposed FCL-DVR is shown in Fig. 1. By comparing the proposed structure with that presented in [31], it is shown that using only two IGBTs a simpler structure is realized in this paper, that can both limit the fault current and compensate the voltage sag. The circuit configuration of this FCL-DVR consists of a transformer, two IGBTs, two reverse diodes and a dc voltage source. Generally, the DC energy source of dynamic voltage restorer (DVR) is selected based on the design and manufacturer of the DVR. Since DVR is for injecting only for a short time, DC capacitors and batteries drawn from the line through a rectifier are frequently used. However, other DC power sources that can be used are supercapacitors, superconducting magnetic storage units, and flywheels. In this case, the $V_{dc}$ of test setup is provided from a 10V DC power supply. The transformer is used for magnetic coupling between the main line and the control circuit or FCL-DVR [33], [34]. The primary side of the transformer is placed in series with the main line. Transformer turn ratios would be adjusted based on the voltage and current stresses on the elements on the secondary side of the transformer and voltage sag that may occur in the system. The capacitor $C_1$ is used as an output filter. Inductor $L_m$ is used as the magnetizing inductance of the transformer and it is considered as the current limiting element in a limiting mode. DC voltage source is used to compensate the voltage of the load side. It should be noted that during voltage restoration function, transformer can experience more flux-linkage two times higher than normal condition. Therefore, in order to avoid transformer saturation, the rating flux should be double that of the steady-state limit. The circuit has three modes of operation as follows:

1) Normal mode
2) Voltage sag compensation mode,
3) Fault current limiting mode.

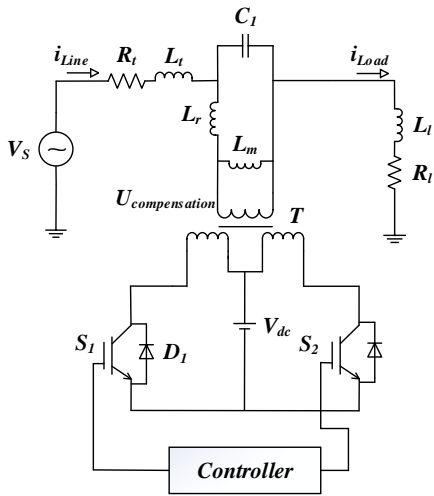

Fig. 1. The proposed FCL–DVR structure.

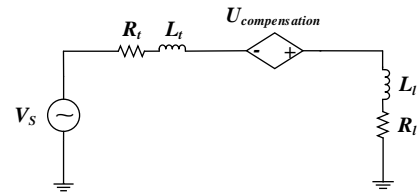

Fig. 2. The equivalent circuit of network in compensation mode.

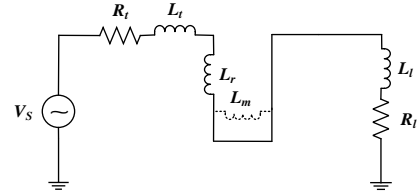

Fig. 3. The equivalent circuit in normal mode.

The abovementioned modes of operation are explained in following parts.

### A. Normal mode

In this mode, the system operates without any problem, and the FCL-DVR bears no effect on the operation of the system.

### B. Voltage compensation mode

The equivalent circuit of this mode is shown in Fig. 2. In the normal operation that voltage compensation is not needed, two switches $S_1$ and $S_2$ are ON, and the DC source is ineffective.

Thus, the secondary of the transformer is short-circuited and will have no effect on the main network and $U_{comp}$ is zero. When $S_1$ and $S_2$ are conducting, magnetizing inductance of the transformer is short-circuited and has no affects on the network. Meanwhile, if a voltage sag occurs, $S_1$ and $S_2$ operate as an inverter and create $U_{comp}$ at the primary side of the transformer and thus compensate the voltage sag. Therefore, the power quality improves.

### C. Fault current limiting mode

When a fault occurs in the network, this mode commences operation. In this mode, turn-off commands are sent to the switches $S_1$ and $S_2$ by the control circuit. Therefore, fault current will pass through $L_m$ and the fault current will be limited. Hence, the fault current is limited by $L_m$. Therefore, contrary to other structures, the fault current has been limited and voltage compensation is performed at once.

## III. THEORETICAL ANALYSIS

### A. Circuit Analysis

In this Section, operation modes of circuit during normal and faulty condition will be discuss. In the first mode, when the network is in a normal state, the switches $S_1$ and $S_2$ are ON. Therefore, the FCL-DVR has no effects on the grid voltage and current. The effect of $C_1$ is negligible since operating frequency is 50Hz. The equivalent circuit of this mode is shown in Fig. 3. The circuit equation for this mode is:

$$V_m \sin\omega t = (L_t + L_l + L_r)\frac{di_L}{d_t} + (R_t + R_l)i_L \quad (1)$$

Solving (1), the line current is obtained as:

$$i_L = \frac{V_m}{Z}\sin(\omega t - \varphi) \quad (2)$$

Where

$$Z = \sqrt{R_{total}^2 + \omega^2 L_{total}^2} \tag{3}$$

$$\varphi = \tan^{-1} \frac{\omega L_{total}}{R_{total}} \tag{4}$$

$$R_{total} = R_S + R_l \tag{5}$$

$$L_{total} = L_t + L_l + L_r \tag{6}$$

In the second mode, when a fault occurs, the current will start to increase. In order to limit the current, the switches $S_1$ and $S_2$ should be turned off. The equivalent circuit of the fault mode is shown in Fig. 4. The line current equation in fault mode is:

$$V_m \sin\omega t = (L_t + L_l + L_m)\frac{di_L}{d_t} + (R_t + R_l)i_L \tag{7}$$

Since $L_m$ is much higher than $L_r$, $L_r$ can be neglected without causing considerable errors. Solving (7), the line current is obtained as:

$$i_L = Ae^{-\frac{R_{total}}{L_{total}}w(t-t_f)} + B\sin(wt - j) \tag{8}$$

$$A = -\frac{V_m}{\sqrt{R_{total}^2 + \omega^2 L_{total}^2}}, B = \frac{V_m}{\sqrt{R_{total}^2 + \omega^2 L_{total}^2}}, \tag{9}$$

$$\varphi = \tan^{-1}\left(\frac{\omega L_{total}}{R_{total}}\right)$$

### B. Voltage Sag

The fault current will cause synchronous disturbance in the network and insulation damage and it will create a voltage sag in PCC. The equivalent circuit of the system during normal operation is shown in fig. 5. From Fig. 5, the voltage at PCC can be written as:

$$V_{PCC_N} = |V_S|\left|\frac{Z_L}{Z_L + Z_S + Z_T}\right| \tag{10}$$

In (10), $Z_L$ is the total line impedance and load impedance. The $Z_S$ and $Z_T$ are impedances of the source and transformer leakage inductance, respectively. Now, if a fault occurs, based on the equivalent circuit of the system shown in Fig. 6, the equation of PCC voltage at fault mode can be written as:

$$V_{PCC_f} = |V_S|\left|\frac{Z_{Line} + Z_{FCL}}{Z_{Line} + Z_S + Z_T + Z_{FCL}}\right| \tag{11}$$

Since $|Z_L| \gg |Z_{FCL}|$, PCC voltage at normal mode, $V_{PCCN}$ is greater than the voltage at fault mode, $V_{PCCF}$. As $V_{PCCN} > V_{PCCF}$, a voltage sag is created in PCC. This voltage sag can be higher if the fault current is not limited. In this case, $|Z_{FCL}|$ is almost zero in (11) and $|V_{PCC}|$ decreases significantly. It necessitates the use of DVR along with the FCL. The proposed structure, which limits the fault current also compensate the voltage sag.

### C. Voltage Stress of Switch

To quantify the rating of switches associated with each of the compared FCL-DVR, total voltage rating of the switches (*TVRS*) is considered that can be a good measure to compare the economic justification of proposed FCL-DVR. The voltage stress of switches is calculated in two modes:

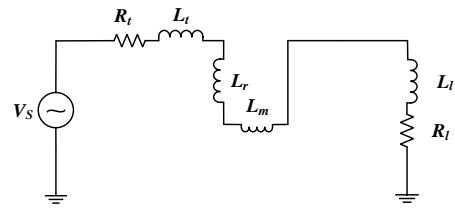

Fig. 4. Equivalent circuit of grid in fault mode.

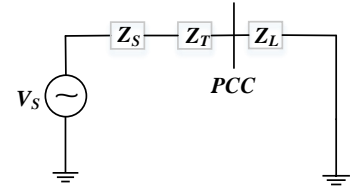

Fig. 5. The equivalent circuit of system during normal operation.

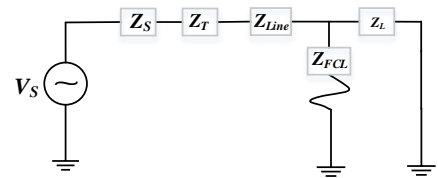

Fig. 6. The equivalent circuit of system during fault operation.

#### 1) Fault Current Mode

In the faulty mode, the switches $S_1$ and $S_2$ endures approximately the same voltage stress, and the voltage stress is defined by the following equation:

$$V_{Stress} = V_{max} \tag{12}$$

The voltage across inductor $L_m$ in the normal mode is zero and when the fault occurs, the switches are turned off and the voltage on $L_m$, can be written from the Fig. 4:

$$V_{Lm} = V_s - V_{Lr} \tag{13}$$

The line current and voltage in normal operation mode are:

$$I_S = I_m \sin(\omega t) \tag{14}$$

$$V_S = V_m \sin(\omega t) \tag{15}$$

When the fault occurs, the switches have high voltage stress so, the voltage stresses are obtained as follows:

$$V_{S_{1max}} = V_{S_{2max}} = \frac{V_{Lm}}{a} \tag{16}$$

$$V_{Lm} = V_S - V_{Lr} \tag{17}$$

$$V_{S_{1max}} = V_{S_{2max}} = \frac{\alpha V_m}{a} + V_{DC} \tag{18}$$

The voltage stress in the half-negative period is:

$$V_{S_{1max}} = V_{S_{2max}} = -\frac{\alpha V_m}{a} + V_{DC} \tag{19}$$

Since the magnitude of $V_{slmax}$ is positive, the fault current cannot pass through the reverse biased diodes. According to the fact that $L_m \gg L_r$, the $V_{Lr}$ can be neglected. Turns ratio of transformer is derived as:

$$V_{Sag} = \lambda V_{Line} \tag{20}$$

$$U_{Comp} \approx V_{Sag} \tag{21}$$

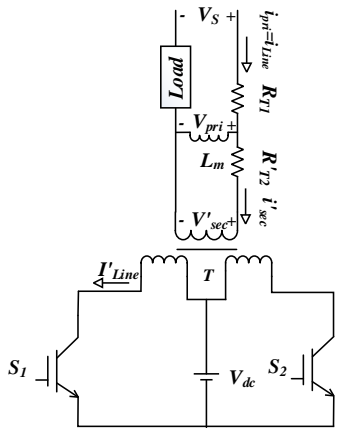

Fig. 7. Detail equivalent circuit of transformer.

$$U_{Comp} = aV_{ac,inv} \quad (22)$$

$$a = \frac{\lambda V_{Line}}{V_{ac,inv}} \quad (23)$$

"a" as turns ratio of the transformer is fixed and should be considered in the primary design of FCL-DVR.

*2) Compensation Mode*

If necessary, the structure will increase the voltage of the transformer to compensate the voltage drop on line. It is important to note that the switches have complementary operation. The voltage stress of switches is calculated as follows:

$$V_{S_{1max}} = V_{S_{2max}} = 2V_{DC} \quad (24)$$

In (22), (23), $V_{ac,inv}$ is the inverter output voltage.

### D. Power losses calculation

In this section, the power loss of the proposed structure is calculated. The equivalent circuit of the transformer is shown in Fig 7. From Fig. 7, the power loss in normal and compensation modes are:

$$\begin{aligned} P_{loss} &= P_{trans} + P_{igbt} \\ &= P_{core} + P_{cu} + V_{igbt}i^*_{Line} = P_{fe} + R_{T1}i^2_{pri} + R^*_{T2}i^{*2}_{sec} + V_{igbt}i^*_{Line} \end{aligned} \quad (25)$$

In faulty mode, the transformer loss is the only part of power loss which can be derived as:

$$P_{loss} = P_{trans} = P_{core} + P_{cu} = P_{fe} + R_{T1}i^2_{pri} + R^*_{T2}i^{*2}_{sec} \quad (26)$$

### E. DC Voltage Selection

When short circuit fault happens, the FCL-DVR will almost completely maintain the DC voltage. Since the voltage sag mainly is not high, therefore, it is not essential to produce high compensation voltage. To ensure the stability of supply voltage, the value of dc link voltage should be higher than the secondary side peak voltage of series and shunt transformers [32]. The dc link voltage is estimated as:

$$V_{DC} \leq 0.65 V_{CES} \quad (27)$$

### F. $L_m$ Design

During fault limiting period, $|Z_T| \gg |Z_L+Z_S|$. Therefore, the fault current can be written as:

$$I_F = \frac{U_S}{Z_T + Z_S + Z_L} \approx \frac{U_S}{Z_T} = \frac{U_S}{\omega L_{m,Max}} \quad (28)$$

Also, the fault current is λ times of the rated load current:

$$I_F = \lambda I_{L,\max} \quad (29)$$

Therefore, $L_m$ can be achieved as:

$$L_{m,Max} = \frac{U_S}{\omega \lambda I_{L,\max}} \quad (30)$$

### G. Series Transformer Design

The capacity and ratio of series transformer should be designed based on maximum fault current limiting capacity. In (29), the fault current is defined as λ times of the rated load current. The rated load current can be calculated as:

$$I_{L,\max} = \frac{S_{Load}}{V_S} \quad (31)$$

The maximum fault current can be achieved based on (28)–(31) as follows:

$$I_{F,\max} = \lambda I_{L,Max} = \frac{U_S}{k^2 \omega I_L} \quad (32)$$

Therefore, the turn ratio of series transformer is derived as:

$$k = \frac{U_S}{\lambda I_{L,Max} \omega I_L} \quad (33)$$

Also, the capacity of series transformer can be achieved as:

$$S_T = \lambda I_{L,Max} U_S \quad (34)$$

## IV. COMPARISON OF THE PROPOSED FCL-DVR WITH OTHER SIMILAR TOPOLOGIES

In Table I, the proposed structure is compared with other available FCLs, DVRs and FLC-DVRs. To this end, there are some important benchmarks which should be considered in the comparison. These benchmarks are number of transformers, number of switches, number of DC sources, power loss and voltage sag compensation capability. To make a fair comparison, it is assumed that all structures are considered in three phase operational mode.

From Table I, the proposed structure has a smaller number of elements and least power loss. It should be noted that there is some configuration without voltage compensation capability with lower number of switches. However, among structures with both sag voltage compensation and current limiting capability, the proposed structure has the lowest number of active switches.

Another important benchmark is the number of transformers in the FCL. This is mainly because of transformer limitations during fault and their high price. The proposed structure needs only one transformer for operation.

The power loss of different structures is derived and given in Table I. As expected, the proposed FCL–DVR has lower loss in comparison with other structures. This is mainly because of lower number of components in the circuit. The structures in [9] and [22] have both the above-mentioned

Table I
COMPARISON WITH OTHER SIMILAR TOPOLOGIES

| Reference | Number of transformers | Number of switches in 3 phases | Number of DC sources in 3 phases | Power losses in 1 phase mode | Compensation voltage sag | Limiting fault current ability |
|---|---|---|---|---|---|---|
| [1] | 2 | 1 | 0 | $P_{loss} = (i_{pri} R_{S1} + i^{*'}_{Sec} R'_{S2}) i_{Pri} + (2V_{DF} + V_{SW} + r_d I_d) I_{DC}$ | NO | YES |
| [23] | 1 | 1 | 1 | $P_{loss} = (i_{pri} R_{S1} + i^{*'}_{Sec} R'_{S2}) i_{Pri} + (\sqrt{2} I_N R_C + 2V_D + V_{igbt}) I_{DC}$ | NO | YES |
| [24] | 1 | 6 | 0 | $P_{loss} = P_{core} + (R_{T1} i^2_{pri} + R^*_{T2_{series}} i^{*2}_{sec}) + V_{igbt} I^*_{line}$ | NO | YES |
| [22] | 2 | 24 | 3 | $P_{loss} = P_{core_{Tseries}} + (R_{T1_{series}} i^2_{pri} + R^*_{T2_{series}} i^{*2}_{sec}) + P_{lossT_{shunt}} + V_{igbt} I^*_{line} + V_{thyristor} I^*_{line}$ | YES | YES |
| [9] | 2 | 10 | 0 | N/A | YES | YES |
| The proposed structure | 1 | 6 | 3 | $P_{loss} = P_{core} + R_{T1} i^2_{pri} + R^*_{T2} i^{*2}_{sec} + V_{igbt} i^*_{Line}$ | YES | YES |

capabilities. However, they have more loss and number of elements than the proposed structure.

## V. SIMULATION RESULTS

Power system computer aided design (PSCAD) /electromagnetic transients including DC (EMTDC) simulation is used to verify the validity of the proposed topology and design methodology. Specifications of the implemented prototype in the simulation are listed in Table II.

To validate the proposed structure, supply voltage, load voltage, load current, and voltage on the primary side of the transformer, are obtained from the simulation. The simulation is performed for two operation modes (fault current limiting mode and voltage sag compensation mode). The circuit is designed to provide 30% voltage sag compensation.

### A. Fault current limiting mode

The fault occurs at 250ms. The fault continues for a period of 100ms. From Figs. 8-11, when the fault occurs, with the use of the control circuit, the load voltage is almost zero and the fault current is limited as well. Therefore, the proposed structure can limit the fault current without any problems. It should be noted that fig. 8 shows instantaneous waveform of current. Therefore, based on the time of fault occurrence, it can experience positive or negative peak current.

### B. Voltage sag compensation

According to Figs. 8-11, between 100ms and 200ms that voltage sag has been created and the voltage is reduced by about 28%. It can be seen, by proper control circuit, the required voltage is created to compensation and $V_{Load}$ remains almost unchanged. The total harmonic distortion of source voltage with and without voltage sag compensation are shown in Figs. 12–13, respectively. It is clear cut that the unwanted harmonics in PCC voltage is reduced considerably.

TABLE II
PARAMETERS OF SIMULATION COMPARISON WITH OTHER SIMILAR TOPOLOGIES

| Symbol | QUANTITY | Value |
|---|---|---|
| $V_S$ | Source Voltage | 220V |
| $R_S$ | Source Resistance and Line Resistance | 0.1Ω |
| $L_S$ | Source Inductance and Line Inductance | 0.5mH |
| $f$ | Power System Frequency | 50Hz |
| $L_r$ | Leakage Inductance | 0.0017H |
| $L_m$ | Magnetizing Inductance | 0.08H |
| $R_{load}$ | Load Resistance | 45Ω |
| $L_{load}$ | Load Inductance | 0.01H |
| $a$ | Isolation Transformer Turn Ratio | 1:5 |
| $V_{dc}$ | DC Source Voltage | 40V |

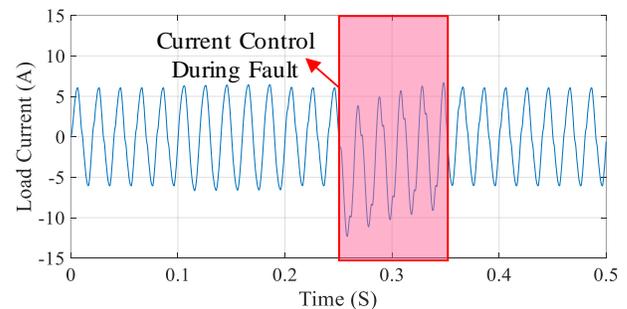

Fig. 8. Simulation waveforms of load current during fault and voltage sag compensation modes.

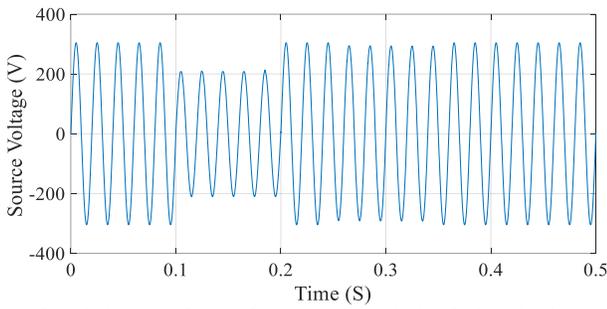

Fig. 9. Simulation waveforms of source voltage during fault and voltage sag compensation modes.

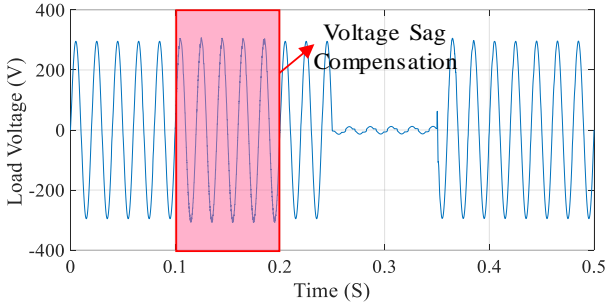

Fig. 10. Simulation waveforms of load voltage during fault and voltage sag compensation modes.

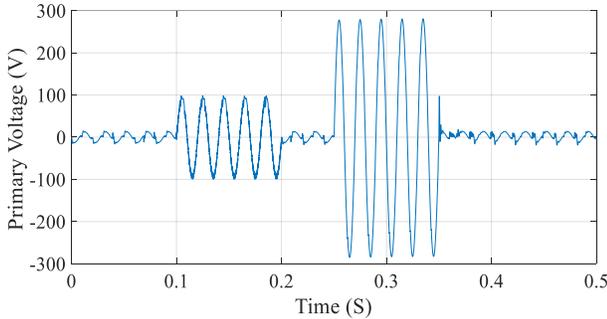

Fig. 11. Simulation waveforms of primary winding voltage of transformer during fault and voltage sag compensation modes.

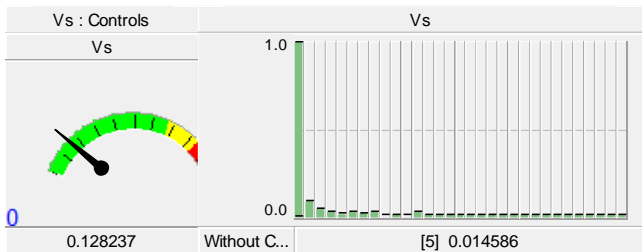

Fig. 12. Total harmonic distortion of source voltage without voltage sag compensation.

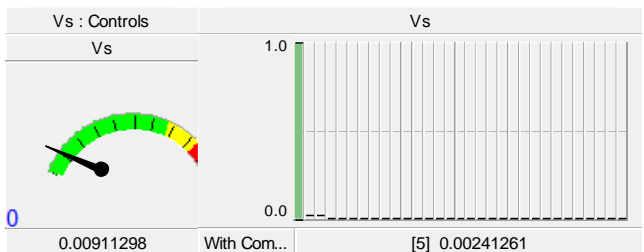

Fig. 13. Total harmonic distortion of source voltage with voltage sag compensation.

## VI. Experimental Results

The proposed structure was developed to prove the feasibility and operation of the circuit. The supply voltage is set to be 63V, the ratio of the transformer is 1:5, and capacitor filter is 20μF. The circuit parameters and manufacturer numbers of the components are given in Table III. In this experiment BUP314 IGBTs are used. Also, the microcontroller in this setup is based on ARM Cortex-M4 core. The Vdc of test setup is provided from a 10V DC power supply. The laboratory testbed for the FLC-DVR is as shown in Fig. 14. Because of limitation of elements in the laboratory, the power rating of the experimental is lower than simulation and listed in Table III.

It should be noted that, the simulation results are presented based on the values in real applications to provide more insights. However, due to the experimental limitations in the lab, the setup is tested at low power range. It is shown that the trajectory of result in both simulation and experimental are confirming each other.

For validation of proposed FCL–DVR, two events are programmed in the experiment. First, there is a high load injection between 0.1S and 0.2S. This event is repetitive and will happen 85ms after fault clearance. In this step, the voltage sag occurs, and it is compensated by the proposed structure.

Fig. 15 (a)-(b) shows before and after compensation. The voltage sag occurs with the depth of about 28%. In Fig. 15 (a)-(b), region in which the waveform becomes zero is related to the fault occurrence zone.

The second event is short circuit between 0.25S and 0.35S. In Fig. 15 (c), after fault occurrence, the fault current limitation strategy is performed to compensate sag voltage. As it can be seen, the proposed structure not only compensates voltage sag but also can limit fault current. In this test setup, the difference between input and output power is about %5 which shows decent efficiency of proposed FCL–DVR. This is mainly because of lower number of active switches during normal operation.

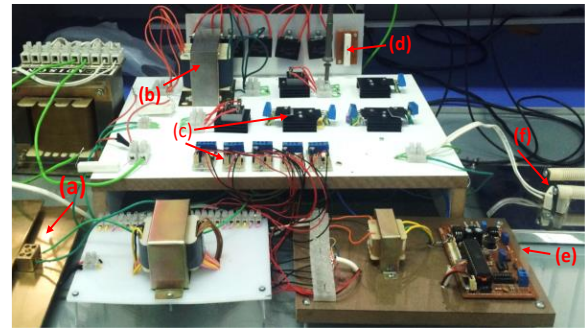

Fig. 14. Setup of experimental prototype. (a) Voltage sag generator, (b) Series transformer, (c) IGBT and gate driver, (d) Filter, (e) Control circuit, (f) Load.

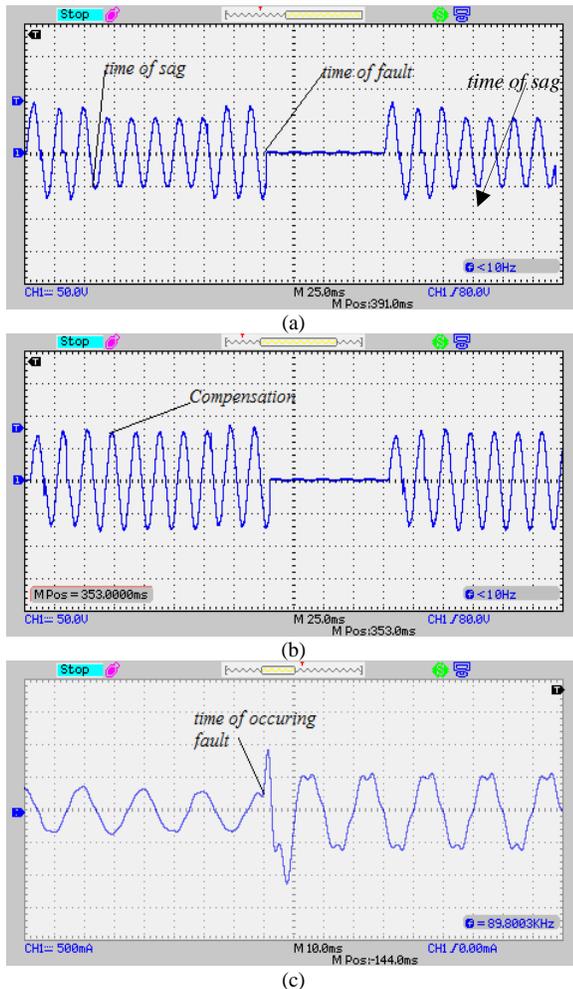

Fig. 15. Experimental result of the proposed FCL-DVR in three modes (a) before compensation [25 mS/div], (b) after compensation [25 mS/div] and (c) limiting fault current [10 mS/div].

TABLE III
PARAMETERS OF EXPERIMENTAL SETUP

| Symbol | QUANTITY | Value |
|---|---|---|
| $V_S$ | Source Voltage | 63V |
| $V_{dc}$ | DC Source Voltage | 10V |
| $a$ | Isolation Transformer Turn Ratio | 1:5 |
| $L_{load}$ | Load Inductance | 0.001H |
| $f$ | Power System Frequency | 50Hz |
| $L_r$ | Leakage Inductance | 0.0017H |
| $L_m$ | Magnetizing Inductance | 0.08H |
| IGBT | Switches | BUP314 |
| $R_{load}$ | Load Resistance | 45Ω |

## VII. CONCLUSION

In this paper, a new fault current limiter–dynamic voltage restorer with reduced number of switches is proposed. The proposed structure has been analyzed and operational modes are explained. The proposed structure utilizes fewer elements in comparison with the structures presented before for fault current limiter-dynamic voltage restorer. This issue directly has effect on cost as the most important factor in industry. On other hand, switches have considerable losses while they have been switching. Hence, the power losses have been reduced. Therefore, the efficiency of the proposed circuit is better in compare to presented FCL–DVR structures in the literature. This claim was verified by theoretically and experimentally. Furthermore, the proposed FCL-DVR not only limits the fault current but also completely compensates the voltage sag and improves power quality simultaneously. The simulation results and experimental results can prove the feasibility of structure in limiting the fault current and improving the voltage quality by compensation for utility voltage sag.

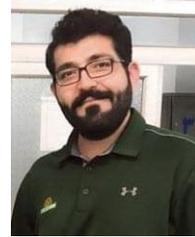

**Pedram Ghavidel** received the B.Sc. degree in electrical engineering and the M.Sc. degree in power engineering from the Department of Electrical Engineering, University of Tabriz, Tabriz, Iran, in 2013 and 2017, respectively.

His research interest includes power electronics, power quality issues, grid transients, fault current limiters, grid-tied converters, and FACTS devices.

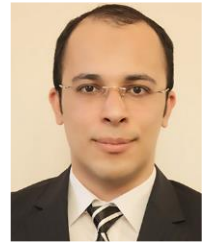

**Masoud Farhadi** (S'20) received the B.Sc. degree in electrical engineering with honors and the M.Sc. degree in power engineering (power electronics and systems) with honors from the Department of Electrical Engineering, University of Tabriz, Tabriz, Iran, in 2013 and 2016, respectively. He is currently working toward the Ph.D. degree at the University of Texas at Dallas, Richardson, TX, USA.

His current research interests include analysis and control of power electronic converters, reliability of power electronic systems, wide bandgap semiconductor device's reliability, and renewable energy conversion systems.

Mr. Farhadi received the 2020 Jonsson School Industrial Advisory Council Fellowship from the University of Texas at Dallas.

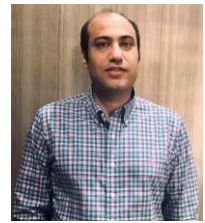

**Morteza Dabbaghjamanesh** (SM'19) received the M.Sc. degree in electrical engineering from Northern Illinois University, DeKalb, IL, USA, in 2014, and the Ph.D. degree in electrical and computer engineering form Louisiana State University, Baton Rouge, LA, USA in 2019. Currently, he is a Research Associate in the Design and Optimization of Energy Systems (DOES) Laboratory at the University of Texas at Dallas, Richardson, TX, USA. His current research interests include power system operation and planning, reliability, resiliency, renewable energy sources, cybersecurity analysis, machine/deep learning, smart grids, and microgrids.

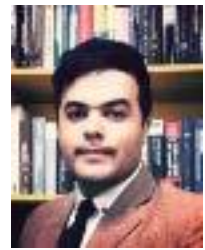

**Alireza Jolfaei** received the Ph.D. degree in Applied Cryptography from Griffith University, Gold Coast, Australia. He is the Program Leader of Master of IT in Cyber Security in the Department of Computing at Macquarie University, Sydney, Australia. His current research areas include Cyber and Cyber-Physical Systems Security. He has authored over 100 peer-reviewed articles on topics related to cybersecurity. He has served over 10 conferences in leadership capacities including program co-Chair, track Chair, session Chair, and Technical Program Committee member, including IEEE Trust-Com and IEEE INFOCOM. He is a Senior Member of the IEEE and a Distinguished Speaker of the ACM on the topic of Cyber Security.

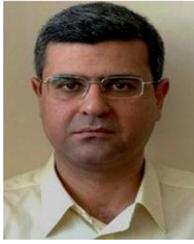

**Mehran Sabahi** was born in Tabriz, Iran, in 1968. He received the B.S. degree in electronic engineering from the University of Tabriz, Tabriz, in 1991, the M.S. degree in electrical engineering from Tehran University, Tehran, Iran, in 1994, and the Ph.D. degree in electrical engineering from the University of Tabriz, in 2009. In 2009, he joined the Faculty of Electrical and Computer Engineering, University of Tabriz, where he has been an Associate Professor since 2014. His current research interests include power electronic converters and renewable energy systems.